\documentclass[unsortedaddress,twocolumn]{revtex4-1}

\usepackage{amsmath}
\usepackage{amsfonts}
\usepackage{amsthm}
\usepackage{amssymb}
\usepackage{graphicx}
\usepackage{hyperref}
\usepackage{color}

	\newcommand{\ncd}{\newcommand}
	\ncd{\mrm}    {\mathrm}
	\ncd{\beq} {\begin{equation}}
	\ncd{\eeq} {\end{equation}}

	\def\d{{\rm d}}

\begin{document}

	\title{Legendre symmetry  and first order phase transitions of homogeneous systems}

	\author{A Bravetti}
		\email{bravetti@icranet.org}
		\affiliation{Dipartimento di Fisica and ICRA, Universit\`a di Roma La Sapienza,\\ P.le Aldo Moro 5, I-00185 Rome, Italy}
		\affiliation{Instituto de Ciencias Nucleares, Universidad Nacional Autonoma de M\'exico,\\
A.P. 70-543, 04510 Mexico D.F., Mexico}

	\author{C S Lopez-Monsalvo}
		\email{cesar.slm@correo.nucleares.unam.mx}
		\affiliation{Instituto de Ciencias Nucleares, Universidad Nacional Autonoma de M\'exico,\\
A.P. 70-543, 04510 Mexico D.F., Mexico}

\author{F Nettel}
		\email{fnettel@ciencias.unam.mx}
		\affiliation{Departamento de F\'isica, Facultad de Ciencias,\\ Universidad Nacional Aut\'onoma de M\'exico, A.P. 50-542, M\'exico D.F. 04510, Mexico}


	\begin{abstract}
	In this work we give a characterisation of first order phase transitions as equilibrium processes on the thermodynamic phase space for which the Legendre symmetry is broken.  
	Furthermore, we consider generalised theories of thermodynamics, where the potential is a homogeneous function of any order $\beta$
	and we propose a (contact) Hamiltonian formulation of equilibrium processes. 
	Indeed we prove that equilibrium corresponds to the zeroth levels of such function. 
	Using these results we infer that the description in equilibrium of first order phase transitions is possible only when the potential is a homogeneous function of order one, unless
	a generalised Zeroth Law is postulated in order to allow for equilibrium between sub-parts of the system at different values of the intensive quantities. Finally, we show the
	 example of the Tolman-Ehrenfest effect.
	\end{abstract}

\maketitle


\section{Introduction}
In the modern treatment of phase transitions it is common to associate the emerging of such phenomena with the breaking of some symmetry  of the system. However, many of the most familiar phase transitions have not been settled yet within this framework, e.g. the liquid-vapour phase transition of water (see for instance the discussion in Sec. 4.1.3 in \cite{Goldenfeld}). Moreover, in  recent years special interest has been devoted to the thermodynamics of non-extensive systems, especially those for which long range interactions play a leading role in their description (c.f. \cite{Tsallis2} and references therein), which cannot be treated in the context of standard extensive thermodynamics and therefore  require a generalisation of the homogeneity condition of the thermodynamic potential.
  
In this work, we consider systems whose thermodynamic potential is a homogeneous function of arbitrary order -- $\beta$ -- of the \emph{extensive} variables, that is
	\beq
	\label{eq.01}
	\Phi(\lambda E^a) = \lambda^\beta \Phi(E^a),
	\eeq
where $\Phi$ represents the thermodynamic potential, $\lambda$ is a constant different from zero, $E^a$ denote the extensive variables and the index $a$ ranges from 1 to $n$, the number of macroscopic degrees of freedom of the system.  We argue that all such systems have a common theoretical symmetry  that breaks whenever a coexistence of two phases is present, i.e. the {\it Legendre symmetry}. To this end, we will adopt a geometric point of view suitable for describing total Legendre transformations as a symmetry acting on the appropriate space. Our results will be consistent with those emerging from the non-equivalence of statistical ensembles as discussed in Chapter 3 of \cite{touchet} and in \cite{Touchette1}. 

 According to the geometric description   introduced by Hermann \cite{Hermann} and Mrugala \cite{mrugala1,mrugala2}, one can define a \emph{thermodynamic phase-space} as a contact manifold where the First Law of Thermodynamics is promoted to an integrability condition defining the integral sub-manifolds of the contact distribution. For a system with $n$ degrees of freedom, the maximal integral sub-manifolds (its {\it Legendre sub-manifolds}) are  precisely $n$-dimensional. This is characteristic of the fact that on such sub-manifolds the thermodynamic potential $\Phi$ and the intensive variables $I_a$  can all be expressed as functions of the $n$ extensive variables $E^a$. 

A symmetry preserving the contact structure of the thermodynamic phase space, induces a diffeomorphism between Legendre sub-manifolds. This is the formal cause of the well known fact that we can use the equations of state to perform a total Legendre transformation, changing the thermodynamic potential and exchanging the role of  the independent variables of the system. We will show that this is always possible as long as the potential satisfies the global convexity conditions, that is, as long as the system is in a single phase \cite{Callen}. 

There are plenty of physical processes whose thermodynamic potentials undergo regions of instability where the Legendre transformation  is not defined. In practice,  one recovers stability by means of the Maxwell equal area law, but the Legendre symmetry cannot be restored \cite{Callen}.  The breaking of the Legendre symmetry allows auto-intersections of Legendre sub-manifolds. This implies that there are processes that leave one thermodynamic phase and go to the other through a sequence of equilibrium states, i.e. the \emph{coexistence processes}. Here, we give a geometric characterisation of such processes as curves lying on the $n-1$ dimensional intersection of two (equilibrium) Legendre sub-manifolds characterising two different phases. 

We introduce a Hamiltonian function on the thermodynamic phase-space which vanishes whenever the system is in equilibrium. A natural candidate for such a function stems from Euler's identity for homogeneous systems, therefore, let us call such function the \emph{Euler's contact Hamiltonian} (ECH). In analogy with Classical Mechanics, all thermodynamic processes are constrained to sub-manifolds where the ECH is constant. Moreover, the exterior derivative of such Hamiltonian is pulled-back to the equilibrium manifold to yield the generalized Gibbs-Duhem relation. We will discuss how this equation imposes constraints on the admissible thermodynamic processes and conclude that equilibrium among the system's sub-parts  requires a reformulation of the Zeroth Law.  These ideas have been discussed in the context of non-extensive thermodynamics (c.f. discussions in  \cite{vanbiro}, and \cite{Tsallis2} for general issues on non-additivity and non-extensivity).  Finally, we discuss an example of a `law' stemming from a generalised notion of equilibrium, i.e. the Tolman-Ehrenfest effect.

\section{Legendre symmetry and phase transitions}

Let us  consider the contact description of the phase space of thermodynamics as given in \cite{mrugala1,mrugala2,CRGTD}.
Given a thermodynamic system with $n$ degrees of freedom, the {\it thermodynamic phase space} is the $(2n+1)$-dimensional  manifold $\mathcal{T}$, 
endowed with a contact structure $\xi\subset T\mathcal{T}$, that is, a maximally non-integrable distribution of co-dimension one hyper-planes.  We can characterise such a distribution with the aid of a 1-form $\theta$ such that 
	\beq
	\xi=\ker(\Theta), 
	\eeq 
and the non-integrability condition
\begin{equation}\label{integraxx}
   \Theta \wedge (\d \Theta)^n \neq 0
\end{equation}
is fulfilled. Equation \eqref{integraxx} can be understood as the necessary condition for a well defined volume form for the thermodynamic phase space. Additionally, it is always possible to find a set of local coordinates for $\mathcal{T}$ such that the 1-form $\Theta$ can be written in the form
	\begin{equation}\label{1stform}
	\Theta = \d \Phi - I_a \d E^a,
	\end{equation}
where we have used Einstein's convention for repeated indices and  $a$ runs from $1$ to $n$ -- the number of degrees of freedom. Note that at this level $\Phi$, $I_a$ and $E^a$ are coordinate functions for $\mathcal{T}$ and do not posses any thermodynamic significance. The thermodynamic nature of such a set of coordinates is realised on the integral sub-manifolds of the thermodynamic phase space. Of special interest are the \emph{maximal} integral sub-manifolds, $\mathcal{E} \subset \mathcal{T}$, i.e. those of maximal dimension which can be embedded in $\mathcal{T}$ such that their tangent bundle is completely contained in the distribution $\xi$. One can show that these correspond to $n$-dimensional sub-manifolds defined through the embedding
	\beq
	\varphi = \left(\Phi(E^a), E^a, I_a(E^b)\right),
	\eeq
satisfying the condition
	\beq
	\varphi^*(\Theta) = \left(\frac{\partial}{\partial E^a} \Phi(E^b) - I_a \right) \d E^a = 0,
	\eeq
equivalent to the First-Law of thermodynamics
	\beq
	\d \Phi(E^a) = I_a \d E^a, \quad \text{where} \quad I_a = \frac{\partial}{\partial E^a} \Phi(E^a).
	\eeq

It is clear that in such sub-manifolds the coordinate $\Phi$ can be interpreted as a thermodynamic potential depending on the \emph{extensive} variables $E^a$, and that their conjugate \emph{intensive} variables $I_a$ correspond to the set of equations of state. Note that the words  {\it extensive} and {\it intensive}  only apply in standard thermodynamics, whereas in the following
we will work with generalised thermodynamics, where the potential and the variables $E^a$ need not be extensive. We will attach ourselves to the use of this convention. Moreover, in the present work  we will only consider situations where the potential $\Phi(E^a)$ is a homogeneous function of order $\beta$.

\begin{figure}
	\centering
	\includegraphics[scale=0.3]{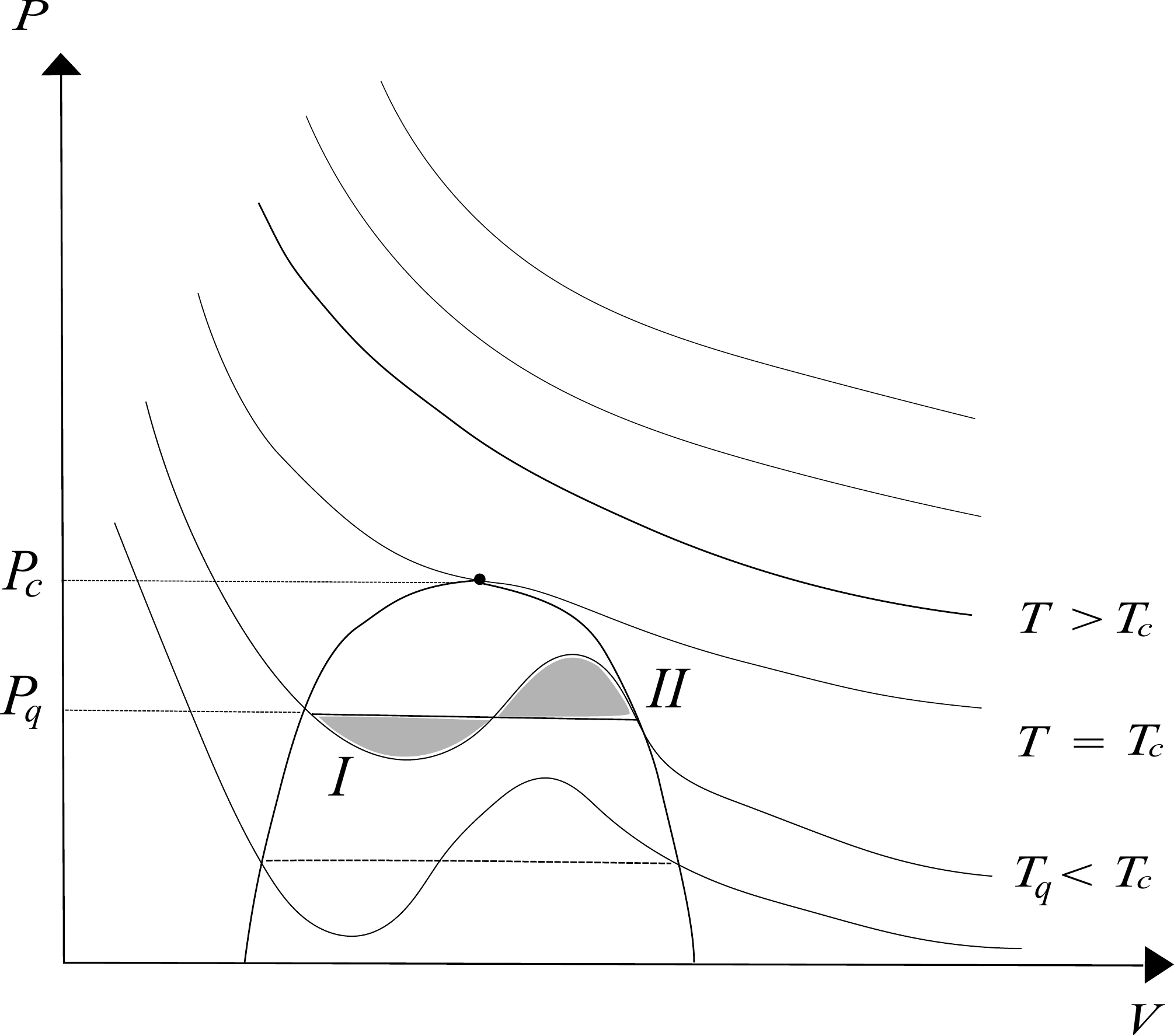}
	\includegraphics[scale=0.32]{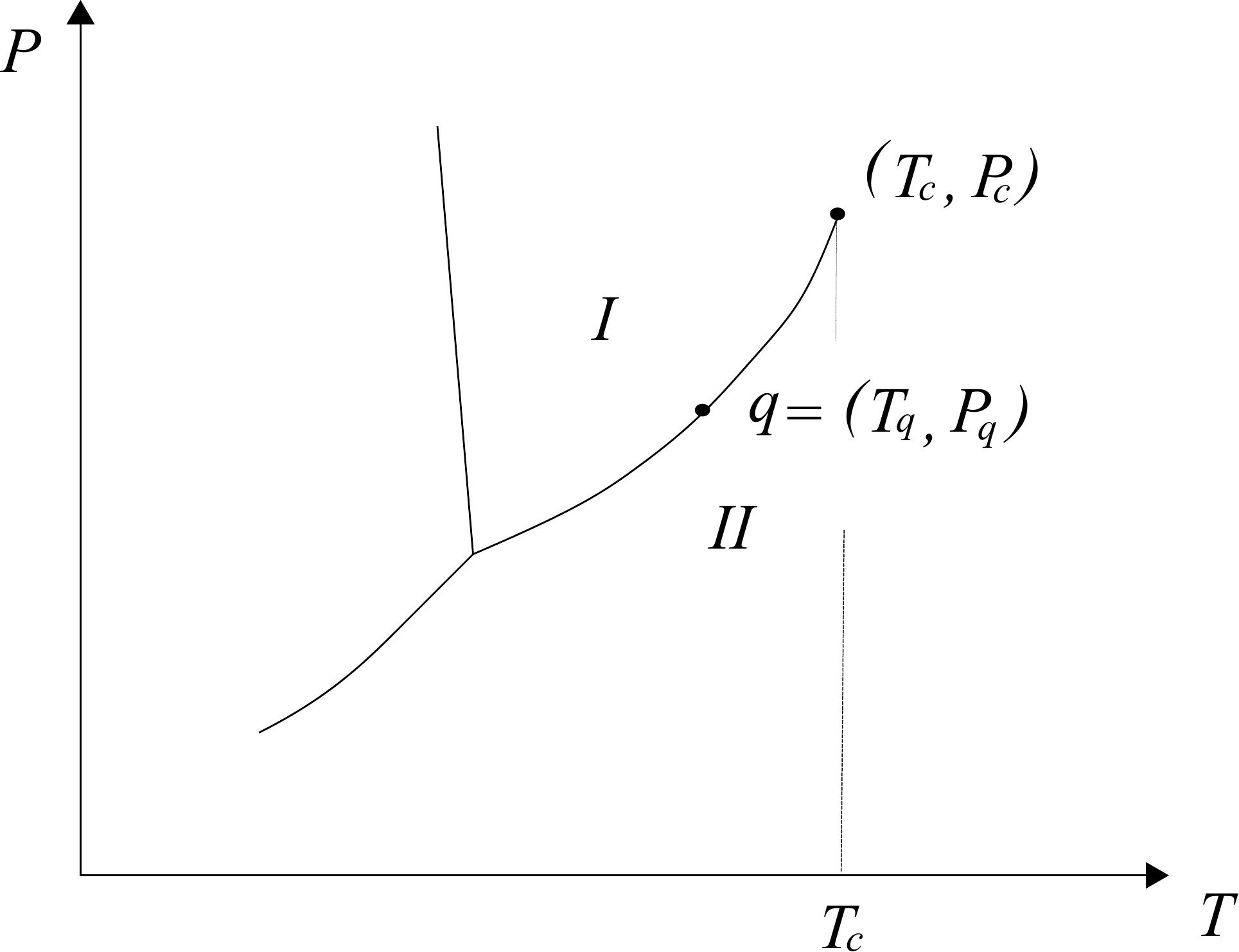}
	\caption{The coexistence process as it appears on a $P-V$ diagram and on a $P-T$ diagram. Two aspects are of major interest. The first is to note that
	the process of coexistence is represented by a line on the $P-V$ diagram and a point on the $P-T$ diagram. Second, in order for the two phases to coexist 
	at equilibrium, the temperatures and the pressures must be the same. More details in the text. }
	\label{figure1}
	\end{figure}

Let us now turn to describe briefly the process of coexistence (for a more detailed description see e.g. \cite{Callen}). To this end, we refer to the $P-V$ and $P-T$ diagrams of the liquid-vapour coexistence for a Van der Waals fluid presented in Fig. \ref{figure1}. Above the critical temperature $T_c$, the isotherms on the $P-V$ diagram are decreasing functions of $V$ and, therefore, are stable. On the contrary, below the critical temperature, the isotherms have a region of instability, which is ``cut-out'' by means of the Maxwell construction, which consists in finding the equilibrium value for the pressure at which the two phases coexist at equilibrium. It turns out that such equilibrium value is given by requiring that the two areas indicated by $I$ and $II$ in the $P-V$ diagram be equal. It is crucial for our discussion to note that the way in which Maxwell construction operates to restore equilibrium is by requiring that the intensive quantities be equal between the two phases \cite{Callen}, which is the standard definition of equilibrium between two parts of a system described by a homogeneous function of order one.  

It is worth noting that when a coexistence of two or more phases is present, the description in the variables $E^a$ is {\it not equivalent} to that  using the variables $I_a$. In fact, by looking at the two diagrams in Fig. \ref{figure1}, one immediately sees that in the $P-V$ coordinates the coexistence is given by a line, whereas in the $P-T$ coordinates it corresponds to the single point, indicated by $q$. As consequence, the change of coordinates in that region is not well defined. From a perspective similar to that of the renormalisation group technique (see e.g. \cite{Kadanoff} and references therein), this signals that in such region new parameters might be involved to account for the (now relevant) extra degrees of freedom, such parameters being in this case the full set of $2n+1$ thermodynamic
variables in the phase space. Furthermore, the need for more degrees of freedom hints at the breaking of some underlying symmetry. Indeed, as we have already pointed out, the descriptions using the extensive or the intensive coordinates are equivalent  \emph{as long as the Legendre transform is well defined}, i.e. when the global convexity conditions are strictly satisfied \cite{Callen}. In fact, the transformation from the thermodynamic potential $\Phi(E^a)$ to its total Legendre transform $\tilde\Phi(I_a)$ induces a diffeomorphism $\psi$ on every Legendre sub-manifold $\mathcal E$ between the coordinates $E^a$ and $I_a$ given by the equations of state 
	\begin{equation}\label{Eos}
	I_a(E^b)=\partial_{E^a}\Phi(E^b).
	\end{equation}
It is straightforward  to calculate the push-forward of such transformation, which is
	\begin{equation}\label{tangentmap}
	\begin{split}
	&\psi_*\,:\,\,\,T\mathcal E\,\,\,\,\,\longrightarrow \,\,\,T\mathcal E\\
	&\psi_*\left(X^a\partial_{E^a}\right)\equiv(\partial_{E^aE^b}\Phi)\,X^b\partial_{I_a}\,,
	\end{split}
	\end{equation}
where $X=X^a\partial_{E^a}$ is any vector field on $T\mathcal E$. For example,  the ideal gas is a system that globally satisfies  the \emph{strict} convexity conditions, therefore, it has only one phase which is represented in contact geometry by a single, smooth, Legendre sub-manifold. However, the majority of systems undergo regions of instability and phase separation. Therefore, the Legendre mapping between the two sets of coordinates is a diffeomorphism {\it as long as the Hessian of the potential $\Phi$ is non-degenerate}. 

There is a subtle point to note here. Whenever we consider a homogeneous  thermodynamic system of order one, the total Legendre transformation is always degenerate, an indication that we are considering more degrees of freedom than needed. Thus, one uses the scaling property of the system and fixes one of the extensive variables and divides the rest of them by such a fixed amount. Since the system is extensive, the result of this also divides the potential by the same amount. In practice, one either chooses the particle number or the volume, and works with molarised or densitisised quantities, respectively, for which the total Legendre transform is well-defined.

Let us explore the consequences of breaking  the Legendre symmetry over a region of coexistence.  Consider the total Legendre transformation in $\mathcal{T}$. This corresponds to a redefinition of the coordinate functions given by
	\begin{align}
	\label{tildepotential}
	\tilde\Phi &	\equiv \Phi-E^a\,I_a\,\\
	\tilde E^a &	\equiv 	- I_a,		\\
	\tilde I_a &	\equiv E^a.
	\end{align}
Such redefinition induces in the equilibrium space $\mathcal{E}$ an exchange of the coordinates $E^a$ for the $I_a$, and now the transformed potential becomes $\tilde \Phi(I_a)$.

First, we look at a thermodynamic  process -- represented by a parametrised curve $\gamma$ --  lying on a subset of $\mathcal{T}$ corresponding to single phase region, that is, $\gamma$ is contained  in a single Legendre sub-manifold $\mathcal E$, where the Legendre transformation is well defined. Along such process, the thermodynamic potential $\Phi(E^a)$ changes according to the first law,  
	\begin{equation}\label{1stlawphi}
	\d \Phi(\dot\gamma)= I_a(\gamma)\d E^a(\dot\gamma),
	\end{equation} 
where we have denoted the tangent vector field to the curve $\gamma$ collectively by $\dot\gamma$. Similarly, its total Legendre transform $\tilde\Phi(I_a)$ also changes according to the first law, this time written as
	\begin{equation}\label{1stlawtilde}
	\d\tilde \Phi(\dot\gamma)= -E^a(\gamma)\d I_a(\dot\gamma)\,.
	\end{equation} 
	
Now, we look at a process taking place in a coexistence region, this time we label it by $\gamma_c$. In this case, the intensive variables are fixed by the requirement of equilibrium between the two phases, while the extensive variables change.  $\Phi$ varies along the coexistence process according to the first law (\ref{1stlawphi}). However,  $\d\tilde\Phi(\dot\gamma_c)=0$, that is, the Legendre transformed potential does not provide us with equivalent thermodynamic information about the process.

The breaking of the correspondence between the changes in the two potentials is a motivation to define on the phase space the \emph{Euler's contact Hamiltonian} (ECH) as 
\begin{equation}\label{ECH}
\mathbb{E}\equiv(1-\beta)\Phi-\tilde\Phi\,,
\end{equation}
where $\tilde \Phi$ is the function defined in (\ref{tildepotential}). Note that this definition can be interpreted as the phase-space function (0-form) generating Euler's identity in the space of equilibrium
 \begin{equation}\label{ECHzero}
 \varphi^*(\mathbb E) = -\beta \Phi(E^a) + I_a\,E^a =  0\,.
 \end{equation}
 From (\ref{ECHzero}) it follows that {\it all the leaves of the contact distribution are characterized by the zeroth level of the ECH}. Also note that, contrary to the symplectic case, every contact Hamiltonian on the thermodynamic phase-space generates a flow tangent only to its zeroth level surface  \cite{mrugala1,mrugala2}. Therefore, the above result implies that the flow generated by the ECH is tangent to the equilibrium sub-manifolds at every point, i.e. it is a thermodynamic process. This suggests that {\it any} thermodynamic process is characterised by the flow of the ECH and completely determined by suitable boundary conditions consistent with the laws of thermodynamics, a matter which will be discussed elsewhere \cite{inpreparation}.

By direct computation we obtain the generalised Gibbs-Duhem relation on each Legendre sub-manifold 
\beq\label{G-D}
\begin{split}
\left.\varphi^*(\d\mathbb E)\right|_{\rm single\,phase}=&\,(1-\beta)\varphi^*(\d\Phi)-\varphi^*(\d\tilde\Phi)=\\
=&\,(1-\beta)I_a\d E^a+E^a\d I_a=0\,.
\end{split}
\eeq
Now, if we consider a process $\gamma_c$ over the region of coexistence (which by the Gibbs phase rule necessarily happens on a sub-manifold whose dimension is not maximal, i.e. it is not a Legendre sub-manifold), one has that $\tilde\Phi$ is constant, that is $\d\tilde\Phi(\dot\gamma_c)=0$, as we have already discussed. Therefore, along a process of coexistence we have that 
\beq\label{G-Dcoex}
\begin{split}
\d \mathbb E(\dot\gamma_c)&=(1-\beta)\d\Phi(\dot\gamma_c)\\
&=(1-\beta)I^{\rm c}_{a}\d E^a(\dot\gamma_c)\,,
\end{split}
\eeq
 where the notation $I^{\rm c}_a$ indicates  that the values of the ``intensive'' quantities are fixed along the coexistence process $\gamma_c$. It is immediate to see that in the case $\beta=1$ the relation (\ref{G-Dcoex}) is identically zero, therefore, both phases are in equilibrium through the coexistence process (having the same values for the intensive quantities) and,  thus, the Gibbs-Duhem equation is satisfied. Nevertheless, when $\beta\neq1$,  the exterior derivative of the ECH -- $\d\mathbb E$ -- along the coexistence is not zero. This implies that the Gibbs-Duhem relation is not identically satisfied. This serves as a motivation to revisit the Zeroth Law of thermodynamics for non-extensive systems (c.f. discussions in \cite{vanbiro} and \cite{oppenheim1}).


\section{Revisiting the Zeroth Law for homogeneous systems}
\label{Oppenheim}

There are numerous equivalent definitions of thermodynamic equilibrium. Here, let us a agree that a system is in equilibrium when an observer cannot distinguish past from future by the sole observation of the macroscopic state of the system at any given time. If this is not the case, the macroscopic state of the system has not settled yet and dissipative fluxes operate in order to `homogenise' the intensive parameters. For example, in the absence of external fields, a system whose temperature is not uniform settles down towards equilibrium by means of heat transfer from regions of higher to lower temperature. This situation is subtler when considering the effect of long range interactions. Thus, for example, one can show that in the case of relativistic heat transfer of a dissipative fluid, the equilibrium condition does not imply that the temperature is uniform through the system's sub-parts, but that the combined effect of the temperature gradient and the external gravitational field (or the fluid's acceleration) balance one another so that there is no net heat flux within the system \cite{eckart,ehlers,tdgr}. This is the well known Tolman-Ehrenfest effect \cite{TolmanEhrenfest}.    
A consequence of such effect is that the entropy becomes  non-extensive.  Generally speaking, all systems with long range interactions have non-extensive properties. A generalised version of the Tolman-Ehrenfest effect for different types of long range interactions has been discussed in \cite{oppenheim1}.

Let us note that our present study independently suggests a different definition for equilibrium in the case of non-extensive systems. In this section we show how to adjust the definition of equilibrium for 
non-extensive homogeneous systems in order to satisfy the generalised Gibbs-Duhem relation, equation (\ref{G-D}), when the system is sub-divided into various parts.

For an ordinary thermodynamic system, the potential $\Phi(E^a)$ is a homogeneous function of order one of the variables $E^a$. In this case,  the conjugate variables $I_a(E^b)$ are homogeneous functions of order \emph{zero}, that is, these are {\it intensive} quantities. This fact is tacitly used in many textbooks to formulate the \emph{Zeroth Law}, namely, that for a given thermodynamic system, the intensive variables attain a  constant equilibrium value through all the system's sub-parts \cite{Callen}. That is, if one scales the system by a non-vanishing constant factor $\lambda$, we have
\begin{equation}\label{hom0}
I_a(\lambda E^b)=\lambda^0 I_a(E^b) = I_a(E^b).
\end{equation}
Following the same reasoning, one can provide a redefinition of the thermodynamic conjugate variables $I_a(E^b)$ to promote them to honest intensive quantities for thermodynamic systems of order different from one. Such redefinition would preserve our primitive notion of equilibrium in homogeneous thermodynamic systems of order one and, moreover, it would allow us to recover the identity character of Gibbs-Duhem relation, that is, it would provide a  consistent definition of first order phase transitions in a more general case.

Let the thermodynamic potential $\Phi(E^a)$ be a  homogeneous function  of order $\beta$ [c.f equation \eqref{eq.01}]. By the properties of homogeneous functions, it follows that the corresponding functions $I_a(E^b)$ are homogeneous of order $\beta-1$ \cite{Belgiorno1}. Thus, we define the generalised intensive quantities by
\begin{equation}\label{Itilde}
\tilde{I}_a(E^b)\equiv\frac{I_a(E^b)}{\left(E^a\right)^{\beta-1}}\,\quad a=1,\dots,n\,.
\end{equation}
It is immediate to verify that the $\tilde I_a(E^b)$ are homogeneous functions of order zero. Moreover, we say that {\it the sub-parts of a system are  in equilibrium between  one another  if and only if the values of the $\tilde I_a(E^b)$ are constant}.

Let  us see now how the Gibbs-Duhem identity is recovered with the new definition of equilibrium over the region of phase separation. First,  recall that in the region where the system is in a single phase, equation (\ref{G-D}) is an identity for any homogeneous function, independent of the specific details in the definition of equilibrium at hand. The problem with the use of non-intensive quantities
 only arises when one divides the system into various parts. Perhaps the most evident -- and the most important --  case is that of coexistence, in which the system naturally separates into two phases. Thus, we start by rewriting  equation (\ref{G-D}) in terms of the $\tilde I_a(E^b)$ defined in (\ref{Itilde}), to obtain
\begin{equation}\label{G-D2}
(1-\beta)\tilde I_a \left(E^a\right)^{\beta-1} \d E^a+E^a \d\big(\tilde I_a \left(E^a\right)^{\beta-1}\big)=0\,.
\end{equation}
Now, considering a process of coexistence $\gamma_c$ as in the preceding section and rewriting equation (\ref{G-D2}) by expanding the second differential, we get
\begin{equation}\label{G-D3}
\begin{split}
&(1-\beta)\tilde I_a \left(E^a\right)^{\beta-1} \d E^a(\dot\gamma_c)+\\&E^a \big[ \left(E^a\right)^{\beta-1}\d \tilde I_a(\dot\gamma_c) + (\beta-1)\left(E^a\right)^{\beta-2}\tilde I_a\d E^a(\dot\gamma_c)\big]=0\,.
\end{split}
\end{equation}
Finally, imposing the condition that the coexistence process happens at equilibrium which,  by our new definition, means that the values of the $\tilde I_a(E^b)(\gamma_c)$ are constant, the first addend in the square brackets vanishes and  we recover an identity.

\subsection{Example: Tolman-Ehrenfest effect}

One of the necessary conditions for thermodynamic equilibrium in general relativity demands the existence of a time-like Killing vector for the space-time under consideration \cite{ehlers}. In the context of a fluid in a spherically symmetric space-time one can show that such a Killing vector satisfies
	\beq
	g_{ab} \kappa^a \kappa^b = g_{00}(r) = -\kappa^2 = - T^{-2},
	\eeq
where $g_{ab}$ are the components of the metric, $\kappa^a$ is the time-like Killing vector, $T$ is the temperature of the fluid and $r$ denotes the radial coordinate. We observe that the condition for thermal equilibrium does not require  the temperature to be constant along the radial direction, but instead it must satisfy $T \left(- g_{00}\right)^{1/2} = \text{constant}$, which in the weak field limit reduces to
	\beq
	T(r) \left(1 + \frac{\phi(r)}{c^2} \right)= \text{constant},
	\eeq
where $\phi(r)$ is the Newtonian gravitational potential and $c$ is the speed of light. Therefore, it follows from \eqref{Itilde} that the entropy of the fluid is
	\beq
	S = \left(1 + \frac{\phi(r)}{c^2} \right)^{\frac{1}{1-\beta}},
	\eeq
and, hence, the system cannot be extensive, i.e. one cannot set $\beta$  equal to 1 unless the gravitational potential identically vanishes.

 
\section{Conclusions}
We have seen that the breaking of the Legendre symmetry can be always associated to first order phase transitions in 
the case of homogeneous thermodynamic systems of any order. Moreover, in the region of coexistence the system separates into two phases and hence new parameters are necessary to fully describe it.
We have argued that such parameters shall be the complete set of thermodynamic variables that constitute the thermodynamic phase space.
Besides, we have introduced a (contact) Hamiltonian energy for thermodynamics that defines all the possible equilibrium processes as orbits 
constrained to sub-manifolds of the thermodynamic phase space defined as zero levels of such energy (a detailed treatment of this will be given elsewhere \cite{inpreparation}).
The construction has been given for homogeneous thermodynamic potentials of any order and therefore these results apply to any homogeneous generalisation of standard extensive thermodynamics. 
 An interesting byproduct of our Hamiltonian formulation of homogeneous thermodynamics is that
 a generalised Zeroth Law of thermodynamics shall be introduced in order to make the notion of equilibrium consistent with the  Gibbs-Duhem relation. This definition can easily be extended to all quasi-homogeneous functions \cite{Belgiorno1} and therefore can apply to all the thermodynamic systems for which some scaling laws are computable. 
 As a particular case, we have used here the Tolman-Ehrenfest effect to show that the internal energy of a thermodynamic system on a gravitational field cannot be homogeneous of order one unless the gravitational potential vanishes. These topics are of much interest, since our investigation is based on a purely mathematical approach. Therefore  we will investigate in more detail the relationship
with the generalisations of the Tolman-Ehrenfest effect and with the equilibrium definition in statistical mechanics of non-extensive systems in a sequent work.

\section*{Acknowledgements}
The authors are thankful to H Quevedo, A C Gutierrez-Pineres and H Touchette for insightful comments and suggestions. 
AB wants to express his gratitude to Instituto de Ciencias Nucleares, UNAM for its kind hospitality during the preparation of this manuscript. 
CSLM acknowledges financial support from CONACYT,  Grant No. 290679\_UNAM. 
FN was supported by DGAPA-UNAM (postdoctoral fellowship).


\end{document}